\pdfoutput=1
\documentclass[11pt,twoside,a4paper,cmspaper,final,collab]{cms-tdr}
\def\svnVersion{6b6064b-D}\def\svnDate{2021/05/05}\def\cmsCernNoTag{CERN-EP-2020-250}\def\cmsCernDate{\today}\def\cmsMessage{"Published in the Journal of High Energy Physics as \href{http://dx.doi.org/10.1007/JHEP05(2021)285}{\doi{10.1007/JHEP05(2021)285}.}"}
\begin{document}\cmsNoteHeader{SMP-19-010}

\newlength\cmsTabSkip\setlength{\cmsTabSkip}{1ex}
\ifthenelse{\boolean{cms@external}}{\providecommand{\cmsLeft}{upper\xspace}}{\providecommand{\cmsLeft}{left\xspace}}
\ifthenelse{\boolean{cms@external}}{\providecommand{\cmsRight}{lower\xspace}}{\providecommand{\cmsRight}{right\xspace}}
\newcommand{\chiso}{\ensuremath{{I}_{\text{ch}}\xspace}}
\newcommand{\ptZ}{\ensuremath{\pt^{\PZ}}\xspace}
\newcommand{\mur}{\ensuremath{\mu_{\textrm{R}}}\xspace}
\newcommand{\muf}{\ensuremath{\mu_{\textrm{F}}}\xspace}
\newcommand{\JETPHOX}{\textsc{JetPhox}\xspace}
\newcommand{\OPENLOOPS}{\textsc{OpenLoops}\xspace}
\newcommand{\zvv}{\ensuremath{\PZ\to\Pgn\Pagn}\xspace}
\newcommand{\wjets}{\ensuremath{\PW + \text{jets}}\xspace}
\newcommand{\zjets}{\ensuremath{\PZ+\text{jets}}\xspace}
\newcommand{\zj}{\ensuremath{\PZ+\text{jets}}\xspace}
\newcommand{\zg}{\ensuremath{\PZ/\gamma}\xspace}
\newcommand{\gj}{\ensuremath{\gamma+\text{jets}}\xspace}
\newcommand{\gjets}{\ensuremath{\gamma+\text{jets}}\xspace}
\newcommand{\dr}{\ensuremath{\Delta R}\xspace}
\newcommand{\mindr}{\ensuremath{{\dr}_{\PZ,\text{j}}}\xspace}
\newcommand{\sigieta}{\ensuremath{\sigma_{\eta \eta}}\xspace}
\newcommand{\mumu}{\ensuremath{m_{\mu \mu}}\xspace}
\providecommand{\cmsTable}[1]{\resizebox{\textwidth}{!}{#1}}

\cmsNoteHeader{SMP-19-010} 
\title
  {Measurements of the differential cross sections of the production of  \texorpdfstring{\zjets}{zjets} and \texorpdfstring{\gj}{gj} and of \texorpdfstring{\PZ}{Z} boson emission collinear with a jet in \texorpdfstring{$\Pp\Pp$}{pp} collisions at \texorpdfstring{$\sqrt{s} = 13\TeV$}{sqrt(s) = 13 TeV}}

\date{\today}

\abstract{Measurements of the differential cross sections of \zjets and \gj production, and their ratio, are presented as a function of the boson transverse momentum. Measurements are also presented of the angular distribution between the \PZ boson and the closest jet. The analysis is based on $\Pp\Pp$ collisions at a center-of-mass energy of 13\TeV corresponding to an integrated luminosity of 35.9\fbinv recorded by the CMS experiment at the LHC. The results, corrected for detector effects, are compared with various theoretical predictions. In general, the predictions at higher orders in perturbation theory show better agreement with the measurements. This work provides the first measurement of the ratio of the differential cross sections of \zjets and \gj production at 13\TeV, as well as the first direct measurement of \PZ bosons emitted collinearly with a jet.}

\hypersetup{
    pdfauthor={CMS Collaboration},
    pdftitle={Measurements of the differential cross sections of the production of  Z+jets and gamma+jets and of Z boson emission collinear with a jet in pp collisions at sqrt(s) = 13 TeV},
    pdfsubject={CMS},
    pdfkeywords={CMS, jet physics, particle correlations and fluctuations},
}

\maketitle 
\section{Introduction}

The production of vector bosons in association with jets in $\Pp\Pp$ collisions provides an important test of the standard model (SM), as well as the opportunity to study major background processes to many searches for physics beyond the SM~\cite{Tricoli:2020vgn}. 
The 13\TeV center-of-mass energy of the CERN LHC and the large integrated luminosity of 36\fbinv, collected in 2016, are used to measure these processes in regions of phase space that were not previously accessible.

This paper presents a measurement of the differential production cross sections of \zjets and \gj, and their ratio for highly energetic bosons. It also provides the first measurement of a \PZ boson produced in close proximity (collinear) to an associated jet. 
Such measurements probe the SM for events with high boson transverse momentum (\pt), and collinear \PZ-jet emission, and provide precision tests of perturbative quantum chromodynamics (QCD) and electroweak (EW) calculations that are implemented in analytical calculations~\cite{Ridder:2015dxa,Boughezal:2015ded} and Monte Carlo (MC) event generators. 
These measurements also provide constraints on parton distribution functions (PDFs)~\cite{Carminati:2012mm,Boughezal:2017nla}, and are relevant in searches for physics beyond the SM, such as dark matter, supersymmetry, and invisible decays of the Higgs boson. The processes of \zjets and \gjets at high boson \pt are key for estimating backgrounds from a \PZ boson decaying to neutrinos (\zvv), whereas the \zg ratio is a theoretical input using \gjets to predict contributions from \zvv~\cite{Sirunyan:2017jix}. An accurate modeling of these processes can improve the potential for discovering new physics.

The differential cross sections for \zjets and \gj can constrain higher-order perturbative QCD and EW calculations that result in a nonnegligible dependence of the cross sections on boson \pt. The EW radiative corrections become large and negative at high energies, because of the presence of Sudakov logarithms that arise from the virtual exchange of soft or collinear massive gauge bosons~\cite{Stirling:2012ak,Denner:2009gj,Denner:2011vu,Denner:2012ts,Denner:2014ina,Kallweit:2015dum}. 
The corrections are logarithmically enhanced at large energies and their impact has been discussed in the context of searches for dark matter~\cite{Lindert:2017olm}, where the dependence of the EW corrections on \pt can lead to effects of the order of tens of percent at large boson \pt. 
Furthermore, developments in theory have led to improved predictions with automated next-to-leading order (NLO) QCD and EW corrections, for instance \SHERPA + \OPENLOOPS~\cite{Schonherr:2016pvn} and \MGvATNLO ~\cite{Frederix:2018nkq}. 
The \zjets and \gj cross sections, and their ratio, at high boson \pt, provide valuable information for probing the magnitude and dependence of these higher-order corrections on boson \pt.
Differential cross section measurements for \zjets and \gj have previously been performed by the ATLAS~\cite{Aaboud:2017hbk,Aaboud:2017cbm,Aaboud:2017kff} and CMS Collaborations~\cite{Sirunyan:2018cpw,Sirunyan:2018gro} at $\sqrt{s} = 13\TeV$. 
A differential measurement of the \zg cross section ratio at $\sqrt{s} = 8\TeV$ has been performed by the CMS Collaboration  using data corresponding to an integrated luminosity of 19.7\fbinv~\cite{Khachatryan:2015ira}. The measurement presented here is the first measurement of this ratio at 13\TeV. 

In contrast to corrections in quantum electrodynamics and QCD, where the massless gauge bosons lead to logarithms that are canceled by the corresponding real-emission corrections, the massive \PW\ and \PZ bosons act as infrared regulators and provide a physical cutoff for the calculations of their cross sections.
The emission of a \PW\ or \PZ boson can contribute significantly to inclusive \wjets and \zjets production at high energies~\cite{Baur:2006sn,Christiansen:2014kba,Krauss:2014yaa}. Such events can be accessed by selecting a high-\pt event topology and studying the region of small angular separation between a \PW\ or \PZ boson and a jet. Measurements of the emission of \PW\ bosons with jets were performed by ATLAS at 8\TeV~\cite{Aaboud:2016ylh} and CMS at 13\TeV~\cite{Sirunyan:2017wgx}. The fully reconstructable decay products from the \PZ boson (in this case decaying to muons) measured in this work, provide a direct measurement of the angular separation between the \PZ boson and the closest jet.

\section{The CMS detector}
The central feature of the CMS apparatus is a superconducting solenoid of 6\unit{m} internal diameter, providing a magnetic field of 3.8\unit{T}. Within the solenoid volume are a silicon pixel and strip tracker, a lead tungstate crystal electromagnetic calorimeter (ECAL), and a brass and scintillator hadron calorimeter (HCAL), each composed of a barrel and two endcap sections.
The ECAL provides coverage in pseudorapidity $\abs{\eta} < 1.48$ in the barrel region (EB), and $1.48 < \abs{\eta} < 3.0$ in two endcap regions (EE). 
Forward calorimeters extend the pseudorapidity coverage provided by the barrel and endcap detectors. Muons are detected in gas-ionization chambers embedded in the steel flux-return yoke outside the solenoid.

Events of interest are selected using a two-tiered trigger system~\cite{Khachatryan:2016bia}. The first level, composed of custom hardware processors, uses information from the calorimeters and muon detectors to select events at a rate of around 100\unit{kHz} within a fixed latency period of less than 4\mus. The second level, known as the high-level trigger, consists of a farm of processors running a version of the full event reconstruction software optimized for fast processing, and reduces the event rate to around 1\unit{kHz} before data storage.

 A more detailed description of the CMS detector, together with a definition of the coordinate system used and the relevant kinematic variables, is reported in Ref.~\cite{Chatrchyan:2008zzk}.

\section{Event simulation}
\label{sec:datasets}
The production of \zjets and the decay to muons is simulated at NLO in QCD using the MC event generator \MGvATNLO (v2.2.2)~\cite{Alwall:2014hca} interfaced with \PYTHIA (v8.212)~\cite{Sjostrand:2014zea} for parton shower (PS) and hadronization. The QCD matrix element (ME) calculation includes up to three final-state partons. The ME-PS matching is performed following the FxFx prescription~\cite{Frederix:2012ps}.  
The cross section of \zjets production for $\ptZ > 50\GeV$, where \ptZ is the transverse momentum of the \PZ boson, is computed at next-to-NLO (NNLO) with \FEWZ (v3.1)~\cite{Melnikov:2006kv}. The renormalization and factorization scales are both set to the sum of the transverse masses of all final state particles and partons.

The \gjets process is generated using the \MGvATNLO generator at both leading order (LO) and NLO in perturbative QCD. For the LO samples, the ME calculation includes up to two final state partons and uses the \kt MLM matching scheme~\cite{Alwall:2007fs} with a matching parameter of 20\GeV to avoid double counting the final states arising from the ME calculations and PS evolution. The NLO samples are generated with up to one parton in the final state and the ME-PS matching is performed following the FxFx prescription. After correcting for the detector effects, the data are also compared with \gjets samples generated at NLO in QCD using the \JETPHOX (v1.3.1)~\cite{Catani:2002ny,Aurenche:2006vj,Belghobsi:2009hx} generator with the Bourhis--Fontannaz--Guillet set II parton-to-photon fragmentation functions~\cite{Bourhis:1997yu}. 
 The choice for the renormalization, factorization, and fragmentation scales in \JETPHOX are all set to the photon \pt: $\mur = \muf = \mu_{\mathrm{f}} =\pt^{\gamma}$. A parton-level isolation criterion is also required by applying a 5\GeV threshold on the transverse energy (\ET), defined as the sum of all parton energies (each multiplied by the sin $\theta$ of their polar angles), around the photon within a cone of radius $\dr = \sqrt{\smash[b]{(\Delta\eta)^2+(\Delta\phi)^2}} = 0.4$, where $\eta$ is the pseudorapidity and $\phi$ is the azimuthal angle. 

The \zjets and \gjets processes are also generated using \SHERPA + \OPENLOOPS (v2.1.0)~\cite{Bothmann:2019yzt} with a matrix element calculation for up to 2 additional partons at NLO in QCD and up to 4 partons at LO in QCD and the approximate NLO EW calculation using the Comix~\cite{Gleisberg:2008fv} and \OPENLOOPS~\cite{Buccioni:2019sur} matrix element generators. This is merged with $\textsc{CSShower}$~\cite{Schumann:2007mg}, the default parton shower in \SHERPA, using the ME-PS matching implemented according to the \MCATNLO method~\cite{Hoeche:2011fd,Frixione:2002ik}. The renormalization and factorization scales are both set to the $\textsc{mets}$ scale setter~\cite{Bothmann:2019yzt}.

The simulation of background processes contributing to the \zjets and \gjets channels are described in the following.

The production of a \PW\ boson in association with jets, where the \PW\ boson decays to a charged lepton and a neutrino, is also simulated with \MGvATNLO and normalized to the cross section calculated at NNLO with \FEWZ.

Top quark pair events are generated with \MGvATNLO and normalized to the inclusive cross section calculated at NNLO together with next-to-next-to-leading logarithmic corrections~\cite{Kidonakis:2012db,Czakon:2013goa}. Single top quark processes are generated at LO with \POWHEG (v2.0)~\cite{Nason:2004rx,Frixione:2007vw,Alioli:2010xd} and normalized to the NLO cross sections for $\PQt\PW$ and \PQt-channel production~\cite{Re:2010bp}, whereas the $s$-channel production is generated at NLO with \MGvATNLO.

{\tolerance=1000 The diboson production processes are generated at NLO as follows: $\PW\PZ$ is generated with \MGvATNLO; $\PZ\PZ$ is generated with a mixture of \POWHEG and \MGvATNLO; $\PW\PW$ is generated with \POWHEG and normalized to the cross section calculated at NNLO; and $\PW\gamma$ and $\PZ\gamma$ are generated with \MGvATNLO.
Finally, multijet QCD events are generated with \PYTHIA at LO.\par}

The {NNPDF}3.0 LO, NLO, and NNLO PDFs~\cite{nnpdf} are used, respectively, with the LO, NLO, and NNLO codes described above. The \PYTHIA program with the CUETP8M1 underlying
event tune~\cite{Khachatryan:2015pea} is used to describe parton
showering and hadronization for all simulated samples.
The full detector response is simulated using the \GEANTfour~\cite{geant} package for all background and signal samples. 

The presence of additional $\Pp\Pp$ interactions in the same or nearby bunch crossings (pileup), corresponding to an average of 23 pileup $\Pp\Pp$ collisions per event in the data, is incorporated into the simulated events. The additional collisions are simulated with \PYTHIA using the NNPDF2.3 PDFs~\cite{Ball:2012cx} and the CUETP8M1 tune.
The simulated event samples are weighted to match the
pileup distribution measured in data. 

\section{Event reconstruction and selection}
\label{sec:selections}

The particle-flow (PF) algorithm~\cite{CMS-PRF-14-001} aims to reconstruct and identify each individual physics-object (PF candidate) in an event, with an optimized combination of information from the various elements of the CMS detector. The energy of photons is obtained from the ECAL measurement. The energy of electrons is determined from a combination of the electron momentum at the primary interaction vertex as measured by the tracker, the energy of the corresponding ECAL cluster, and the energy sum of all bremsstrahlung photons spatially compatible with originating from the electron track. The momentum of muons is obtained from the curvature of the corresponding track. The energy of charged hadrons is determined from a combination of their momentum measured in the tracker and the matching ECAL and HCAL energy deposits, corrected for zero-suppression effects and for the response function of the calorimeters to hadronic showers. Finally, the energy of neutral hadrons is obtained from the corresponding corrected ECAL and HCAL energies.

In the barrel section of the ECAL, an energy resolution of about 1\% is achieved for unconverted or late-converting photons in the tens of GeV energy range. The remaining barrel photons have a resolution of about 1.3\% up to $\abs{\eta} = 1$, increasing to about 2.5\% at $\abs{\eta} = 1.4$. In the endcaps, the resolution of unconverted or late-converting photons is about 2.5\%, whereas the remaining endcap photons have a resolution between 3 and 4\%~\cite{Khachatryan:2015iwa,Sirunyan:2020ycc}.

For each event, hadronic jets are clustered from PF candidates using the infrared- and collinear-safe anti-\kt algorithm~\cite{Cacciari:2008gp, Cacciari:2011ma} with a distance parameter of $\dr = 0.4$. Jet momentum is determined as the vectorial sum of all particle candidate momenta in the jet, and is found from simulation to be, on average, within 5 to 10\% of the true momentum over the entire \pt spectrum and detector acceptance. Pileup can contribute additional tracks and calorimetric energy depositions to the jet momentum. To mitigate this effect, tracks identified as originating from pileup vertices are discarded and an offset is applied to correct for the remaining contributions~\cite{Cacciari:2007fd,collaboration_2011}. Jet energy corrections are derived from simulation studies so that the average measured response of jets becomes identical to that of particle-level jets. Measurements of the momentum balance in dijet, $\gamma + \text{jet}$, $\PZ + \text{jet}$, and multijet events are used to correct for any residual differences in jet energy scale (JES) in data and simulation~\cite{Khachatryan:2016kdb}. The jet energy resolution (JER) amounts typically to 15\% at 10\GeV, 8\% at 100\GeV, and 4\% at 1\TeV~\cite{Khachatryan:2016kdb}. Additional selection criteria are applied to each event to remove jets potentially dominated by anomalous contributions from various subdetector components or reconstruction failures.

The reconstructed vertex with the largest value of summed physics-object $\pt^2$ is the primary $\Pp\Pp$ interaction vertex. The physics objects are the jets, clustered using the jet finding algorithm~\cite{Cacciari:2008gp,Cacciari:2011ma} with the tracks assigned to the vertex as inputs. 

{\tolerance=800
 Muons are measured in the range $\abs{\eta} < 2.4$, with detection planes using three technologies: drift tubes, cathode strip chambers, and resistive-plate chambers. Matching muons to tracks measured in the silicon tracker results in a relative transverse momentum resolution of 1 ($<$7)\%~\cite{Sirunyan:2018} in the barrel and 3\% in the endcaps, for muons with \pt up to 100 (1000)\GeV.
 \par}

The \PZ bosons are identified by their decay into $\mu^{+}\mu^{-}$ pairs. Events for the \zjets analysis are selected online using a high-level trigger that requires a loosely isolated muon with a minimum \pt threshold of 24\GeV. 
Offline, the muon candidates are required to be: reconstructed in the fiducial region $\abs{\eta} < 2.4$; separated from any jets in the event by a distance of $\dr > 0.5$; and isolated, where the isolation is calculated from the sum of the scalar \pt of all PF candidates within an isolation cone with radius $\dr = 0.4$, which is required to be $<$15\% of the muon \pt. Two isolated muons of opposite electric charges are selected. The dimuon invariant mass \mumu is required to be compatible with the \PZ boson mass, in the range of $71 < \mumu < 111\GeV$. \zjets events thus contain \PZ boson and off-peak Drell--Yan + jets production. In case more than one pair is selected, the highest \pt pair is chosen. The muons are each required to have $\pt > 30\GeV$. In addition, to match the photon requirement in the differential cross section ratio of $\PZ/\gamma$, the dimuon system is required to have $\pt >  200\GeV$ and a rapidity in the range $\abs{y} < 1.4$.

Photon events are selected online with a trigger that requires at least one ECAL cluster with $\ET > 175\GeV$, and the ratio of energy deposited in the HCAL to that in the ECAL to be less than 0.15 (0.10) in the EB (EE) region. Offline, photon candidates are required: (i)~to have $\pt > 200\GeV$ and $\abs{\eta} < 1.4$ to ensure the trigger is fully efficient; (ii)~to pass a set of cut-based high-quality identification criteria based on the shape of the electromagnetic shower in the ECAL; and (iii)~to have an isolation energy calculated from all PF candidates (charged hadrons, neutral hadrons, and photons) and corrected for pileup on an event-by-event basis within a cone of radius $\dr = 0.3$~\cite{Khachatryan:2015iwa,Sirunyan:2020ycc}. The photon isolation from charged hadrons, neutral hadrons, and photons is required to be less than: 0.202, $0.264 + 0.0148\pt + 0.000017\pt^{2}$, and $2.362 + 0.0047\pt$, respectively, where the \pt is the photon \pt~\cite{Khachatryan:2015iwa,Sirunyan:2020ycc}.

Jets are required to have $\pt > 40\GeV$ and $\abs{\eta}<2.4$. For both the \zjets and \gjets channels, at least one jet is required to have $\pt > 100\GeV$.

The definitions of the fiducial region in data and simulation for the \zj and \gj selections closely follow the analysis requirements for the reconstructed objects. The \zj selection requires the presence of two muons with opposite electric charges. The muons are each required to have $\pt > 30\GeV$ and $\abs{\eta} < 2.4$, with an invariant mass in the range  $71 < \mumu < 111\GeV$. For simulation, the muon four-vectors have been summed with all the generated photons and leptons within a cone of $\dr = 0.1$. 
Both channels require the vector boson to have $\pt > 200\GeV$, $\abs{\eta}$ or $\abs{y} < 1.4$, and at least one jet with $\pt > 100\GeV$, where the jets are required to be separated from the muons or the photon by a distance of $\dr > 0.5$.

The selection of events for the collinear \PZ boson emission analysis follows that of the \zj region, except that the requirements on the boson \pt and $y$ are removed and instead the threshold on the leading jet \pt is raised to 300\GeV. The distribution of the angular separation between the \PZ boson and the closest jet (\mindr) from data is compared with theoretical predictions for two thresholds on the leading jet \pt: 300 and 500\GeV. The region $\mindr > 2.5$ is dominated by the back-to-back production of a \PZ boson and a jet, while $\mindr < 2.5$ is enhanced in the collinear production. 
The fiducial region for this analysis is defined by the presence of a \PZ boson reconstructed from muons with $\pt > 30$\GeV, $\abs{\eta} < 2.4$, and one jet with $\abs{\eta}<2.4$ and \pt above 300 or 500\GeV.

\section{Background estimation}
\label{sec:Backgrounds}

\subsection{The \texorpdfstring{\zj}{Z+jets} channel}
\label{sec:Zbkg}

The selection of \zj events produces a relatively pure sample of \PZ bosons decaying to muons. Contributions from background processes in the fiducial region are estimated from simulation and subtracted in the results. 
The dominant contributions are from diboson events and vector boson fusion \zj production, which is treated as a background in this analysis. These backgrounds contribute at the level of 2.5 and 1.5\%, respectively, for the \zj analysis and 3.2 and 3.1\% for the collinear \PZ analysis with leading jet \pt above 300\GeV. 
The systematic uncertainty in the MC prediction is a quadratic sum of the uncertainties in the modeling of the muon selection efficiencies, simulation-based systematic uncertainties, uncertainty in the cross section, and the statistical uncertainty due to the limited number of MC events. The dominant uncertainty for boson \pt above 500\GeV comes from the limited number of simulation events, whereas the uncertainty in the cross section dominates at \pt below 500\GeV. For the collinear \PZ analysis, the statistical uncertainty dominates for high \mindr above 3.4 and the very low \mindr below 0.5, whereas the uncertainty in the cross section is dominant for the \mindr region between 0.5 and 3.4.

\subsection{The \texorpdfstring{\gjets}{gamma+jets} channel}
\label{sec:purity}

The largest background contribution to the \gjets region is from QCD multijet processes in which an electron or hadron from a jet is misidentified as a photon candidate. The total background contribution, which is mostly from such misidentified photon events, is estimated from the purity of \gjets events, defined as the fraction of isolated photons from the hard scattering over the number of all photon candidates after the full selection criteria is applied. A template fit method is used to extract a value for the photon purity in each \pt bin, by fitting the data with a sum of the signal and background templates, where the signal denotes the distribution from genuine photons and the background is the distribution from misidentified photons.
The number of isolated photons emitted in the hard scattering is extracted from a fit to the shower shape variable \sigieta, which defines the extent of the shower along the $\eta$ direction within a $5{\times}5$ array of ECAL crystals~\cite{Khachatryan:2015iwa}. This variable provides discrimination between signal and background, owing to the respective narrow and wide lateral spreads in the showers observed from genuine and misidentified photons.

The signal template is obtained from simulated \gj events generated at NLO using \MGvATNLO, selecting all candidates passing the analysis selection criteria and matched to a particle-level isolated photon coming from the hard scattering. The particle-level photon is defined as a prompt photon with the scalar \pt sum of all additional generated stable particles (within a cone of $\dr = 0.4$ around the photon) required to be less than 5\GeV.
The uncertainty in the shape of the signal template is estimated using an alternative MC simulation. A comparison is made between the signal \sigieta distribution from the LO \SHERPA prediction and the nominal distribution from \MGvATNLO. The shape of the \sigieta distribution from the two simulations are similar. The signal template obtained from \SHERPA is provided as an alternative template to the fit.

A data region enriched in misidentified photons is selected using the isolation of the photon candidate, as determined by summing the transverse momenta of only charged hadrons (charged-hadron isolation \chiso). This region is used to obtain the background template; the presence of genuine photons can lead to a background template that looks like the signal distribution and skews the estimate of the purity. The value of \chiso\ used to define this background-enriched region is thus chosen following an optimization to reduce the contamination from signal events and, at the same time, provide a statistically sufficient sample of misidentified photons. The charged-hadron isolation is required to be in the range 10--15\GeV, following an optimization procedure described below. 
Any remaining residual contribution from genuine photons, which is under 6\% for $\sigieta < 0.010$, is subtracted using simulation. Since this subtraction procedure relies on an accurate normalization of the MC prediction in this region of phase space, an additional cross-check is performed to validate it. The definition of the background-enriched region is varied, and the effect of this variation on the \sigieta distribution is studied. The difference in the shape of this distribution in the nominal region and one with a different \chiso\ range is estimated by constructing a quantity $R^{\sigieta}$, defined as the ratio of the number of events in the signal region ($\sigieta < 0.010$) and background-dominated regions ($\sigieta > 0.014$); the latter being dominated by misidentified photons. The behaviour of $R^{\sigieta}$ is studied for a large number of possible background-enriched regions defined within the \chiso\ range of 1--20\GeV. The ratio is most stable when the lower threshold on the \chiso\ is sufficiently far away from the signal region that the contamination from signal events has a small effect on the ratio. The optimal background-enriched region is found to be $10 < \chiso <  15$\GeV. A systematic uncertainty from the choice of this region is determined from the maximum variation of the background template across all possible regions that produce a value for $R^{\sigieta}$ that does not vary with the amount of signal contamination. This shape difference is small, within the statistical uncertainty associated with the number of events in the sideband region. 

\begin{figure}[h!]
  \centering
  \includegraphics[width=0.49\textwidth]{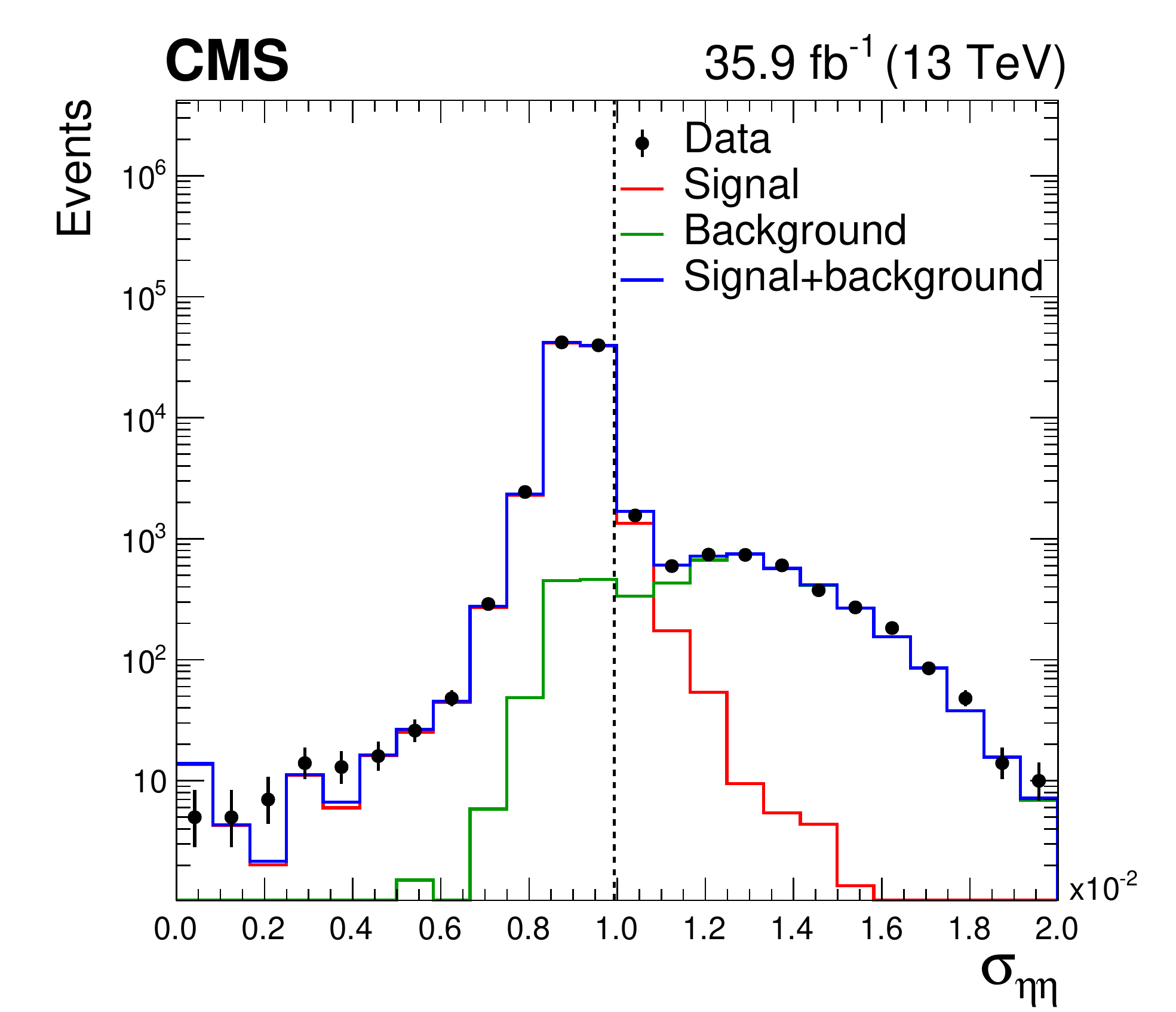}
  \includegraphics[width=0.49\textwidth]{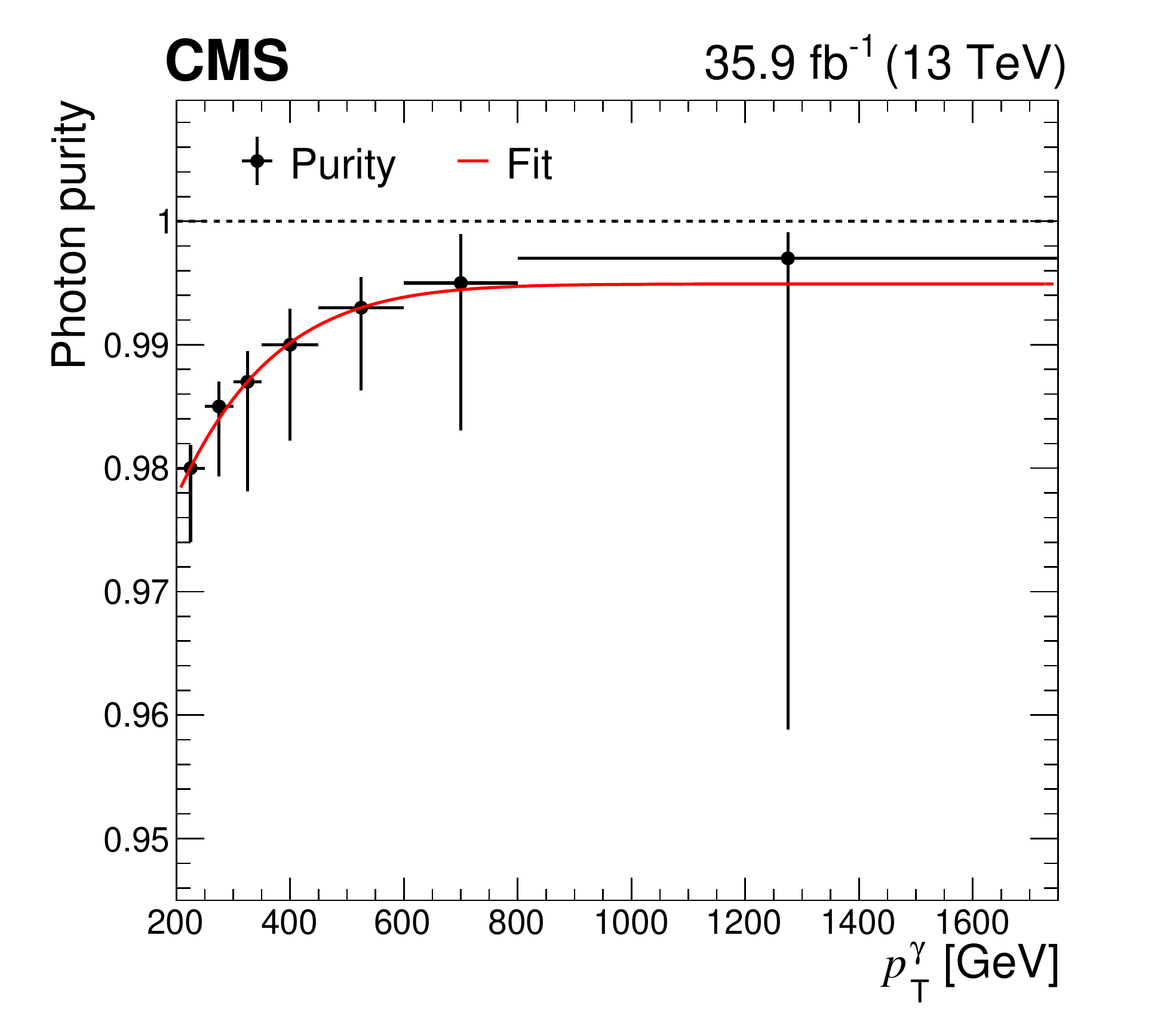}
  \caption{A fit to the \sigieta distribution using signal and background templates to extract a value for the purity in the photon \pt bin of 300--350\GeV. The signal region is to the left of the vertical dashed line (\cmsLeft). The purity, as a function of the photon \pt, is extracted from a fit to the \sigieta distribution in each \pt bin. The error bars show the total statistical and systematic uncertainty and the solid line is the fit to the data points (\cmsRight).}
  \label{fig:purity}
\end{figure}

A binned maximum-likelihood fit is performed to the \sigieta distribution in data to extract the fraction of genuine photons. Statistical uncertainties in the templates are included in the fit as nuisance parameters using the Barlow--Beeston approach~\cite{Barlow:1993dm}.
The purity fraction is estimated by integrating the fitted template over the photon \sigieta fiducial region of $\sigieta < 0.010$. Figure~\ref{fig:purity} shows the results of the fit to the \sigieta distribution for the \pt bin 300--350\GeV, and the purity values extracted from a similar fit in each \pt bin. The purity as a function of photon \pt is then fitted with a functional form and used to extract the purity values for the subsequent unfolding procedure.
Also shown in Fig.~\ref{fig:purity} is the associated uncertainty in the purity, including both the statistical uncertainty from the fit and systematic contributions from the alternative signal template, choice of background-enriched region, discrepancies in modeling of the \sigieta distribution, and photon selection efficiencies. 
The purity for photons with $\pt > 200\, (400)\GeV$ is above 98 (99)\% and approaches 100\% at high \pt.

\section{Corrections for detector effects}

The reconstructed distributions are corrected for the event selection efficiency and detector resolution effects using an unfolding technique that employs a response matrix to map the reconstructed observables onto the generator-level values. 
This is performed using the $\textsc{TUnfold}$ software package~\cite{Schmitt:2012kp}, which is based on a least squares fit and includes the option for a possible Tikhonov regularization term~\cite{Tikhonov:1963}. 
The strength of the regularization parameter is determined with the $L$-curve scan method~\cite{Hansen00thel-curve} and is negligible. Hence, no regularization is applied to the distributions. 

The simulation used to build the response matrix correcting for \zj and \gj events is based on NLO \MGvATNLO.
To build the response matrix, the bosons in each generator-level event passing the fiducial requirements described in Section~\ref{sec:selections} are matched to the corresponding reconstruction-level objects. When the generator-level bosons match the reconstruction-level objects, the response matrix is populated with both events, whereas generated events in the fiducial region without a matching reconstructed boson candidate are used to determine the selection efficiency. Conversely, a reconstructed boson candidate in the fiducial region not matched to a generator-level boson is considered as a further background source.
 
{\tolerance=10000 The unfolding of the \mindr\ distribution is also performed using the NLO \MGvATNLO simulation. For this distribution, an additional matching procedure is performed between the closest jet to the reconstructed \PZ boson and a generator-level jet. The response matrix is then built similarly to the \zj and \gj cases.\par} 

After the unfolding procedure, the fiducial region event yield is obtained and the corresponding measured fiducial cross sections are compared with different theoretical predictions.

\section{Systematic uncertainties}
\label{sec:systematics}
Systematic uncertainties associated with the measurement of the cross sections are propagated by varying the parameter representing the source of each uncertainty by one standard deviation up and down ($\pm 1 \sigma$) and recomputing the response matrix for each variation. 

The systematic uncertainty in the efficiency of selecting muons or photons is determined by comparing the efficiency expected from simulation and measured in data with the ``tag-and-probe'' method~\cite{CMS:2011aa} as a function of the \pt and $\eta$ of the relevant object. For photons, this is derived using a sample of electrons from \PZ decays~\cite{Khachatryan:2015iwa}, reconstructed as photons without the electron-veto requirement. 
This uncertainty is applied as a scale factor on an event-by-event basis and implemented in the unfolding procedure as alternative MC distributions for each $\pm 1 \sigma$ variation on the efficiencies. The statistical (systematic) component of the uncertainty is treated as uncorrelated (correlated) across the boson \pt bins. 
The uncertainty in the muon selection efficiency, which includes the identification, isolation and tracking efficiency, contributes at the 0.8--1.0\% level to the \zj cross section, whereas the uncertainty in the photon selection efficiency, which includes the identification and isolation efficiency, contributes at the 2.5--2.6\% level on the \gj cross section across the full \pt range. The uncertainty in the photon trigger efficiency contributes 0.2--2.2\%, whereas the uncertainty in the muon trigger efficiency is negligible for the \zjets channel and less than 0.2\% for the \PZ-jet collinear region.

The systematic uncertainty in the muon momentum scale is the dominant systematic uncertainty in the \zjets cross section. The uncertainty in the scale of up to 1 (7)\% at $\pt < 100$ (1000)\GeV for each muon results in an uncertainty in the cross section ranging from 1.7\% at low \pt up to 22\% at high \pt.
The uncertainty in the photon energy scale and resolution of 1--2\% results in an uncertainty in the cross section of ($\leq$1\%) at low \pt and up to 8.6\% at high \pt, becoming the dominant systematic uncertainty. 

The effect on the measurement from the uncertainty in the JES and JER is evaluated by varying the jet four-momenta using the uncertainties in the correction factors that depend on the jet \pt and $\eta$ for the JES, and jet $\eta$ for the JER. The uncertainties from the JES and JER are subdominant (below or at the 1\% level) for all three event categories, because of the high-\pt threshold of the leading jet. 

The sources of systematic uncertainty associated with the estimation of the photon purity are described in more detail in Section~\ref{sec:purity}, and contribute up to 1.1\% at low photon \pt and down to 0.2\% at high \pt.

A correction is applied to account for the difference in pileup between data and simulation. 
It has a negligible (less than 1\%) effect on the \zjets and \gjets cross sections.  
A 2.5\% uncertainty in the total integrated luminosity~\cite{CMS-PAS-LUM-17-001} is applied to the unfolded data distribution.

Uncertainties are included in the unfolding procedure from the statistical size of the simulation sample used to build the response matrix and from the difference between the LO and NLO \MGvATNLO samples.
The overall unfolding uncertainty is the quadratic sum of these two contributions, with the dominant uncertainty being the statistical size of the simulation samples.
The uncertainty in the unfolding is the dominant uncertainty at high boson \pt for the \zjets cross section, contributing up to 19\% in the highest \PZ boson \pt bin, and a lower uncertainty in the \gjets cross section, contributing up to 6.4\% at high photon \pt. 
The uncertainty in the unfolding for the \mindr distribution is among the largest uncertainties. 

A summary of the contributions from each uncertainty source to the differential cross section measurements of \zjets, \gjets, \zg ratio, and \PZ-jet angular separation for a leading jet \pt threshold of 300\GeV is shown in Table~\ref{tab:systematics}. Common sources of systematic uncertainties such as those from JES, JER, and integrated luminosity are treated as correlated between \zjets and \gjets and therefore mostly cancel in the \zg ratio, whereas sources of uncertainty such as the lepton efficiency, trigger, and photon purity are treated as uncorrelated.

\begin{table}[h!]
\centering
\topcaption{The contributions to the uncertainty in the differential cross section measurements for the \zjets and \gjets processes, the \zg ratio, and the \mindr region. The uncertainties are expressed in percent, and a range represents the minimum and maximum effect observed.}
\cmsTable{
  \begin{tabular}{lcccc}
    Systematic source & \zjets [\%] & \gjets [\%]& (\zjets)/(\gjets) [\%] & \mindr region [\%] \\ \hline
    Trigger                             & 0.0 & 0.2--2.2 & 0.2--2.2 & 0.0--0.2  \\
    Muon reconstruction and selection   & 0.8--1.0 & \NA         & 0.8--1.0 &0.9--1.1   \\
    Photon reconstruction and selection & \NA         & 2.5--2.6       & 2.5--2.6       & \NA          \\
    Photon energy scale                 & \NA         & 0.5--8.6 & 0.5--8.6 & \NA       \\
    Muon momentum scale                 & 1.7--22 & \NA         & 1.7--22 & 0.1--12  \\
    Photon purity                       & \NA         & 0.2--1.1 & 0.2--1.1 & \NA         \\
    Background yields                   & 0.7--1.5 & \NA         & 0.5--1.6 & 0.9--11 \\
    Pileup                              & 0.0--0.7 & 0.0--0.3 & 0.0--0.4 & 0.2--0.9  \\
    Integrated luminosity                          & 2.5         & 2.5         & 0.0  & 2.5          \\
    Unfolding                           & 0.3--19 & 1.1--6.4 & 1.1--20 & 1.2--11 \\
    JES/JER         & 0.0--0.2 & 0.0--0.2 & $\leq$0.04       & 0.3--1.5  \\ [\cmsTabSkip]
    Total                               & 3.3--29 & 4.0--12 & 4.4--31 & 3.5--17 \\
  \end{tabular}
}
\label{tab:systematics}
\end{table}

\section{Results}

\begin{figure}[!ht]
\centering
  \includegraphics[width=0.49\textwidth]{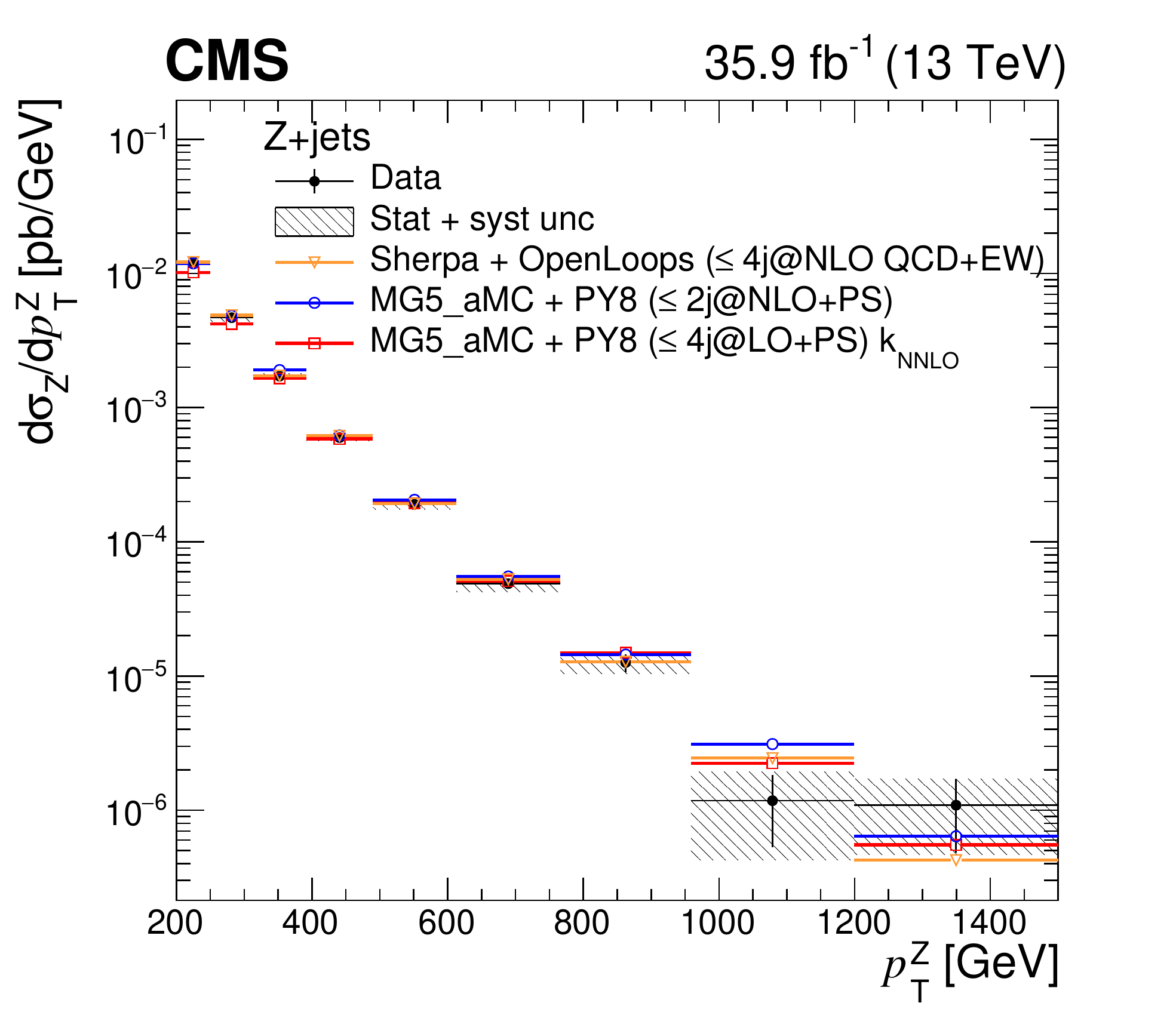}
  \includegraphics[width=0.49\textwidth]{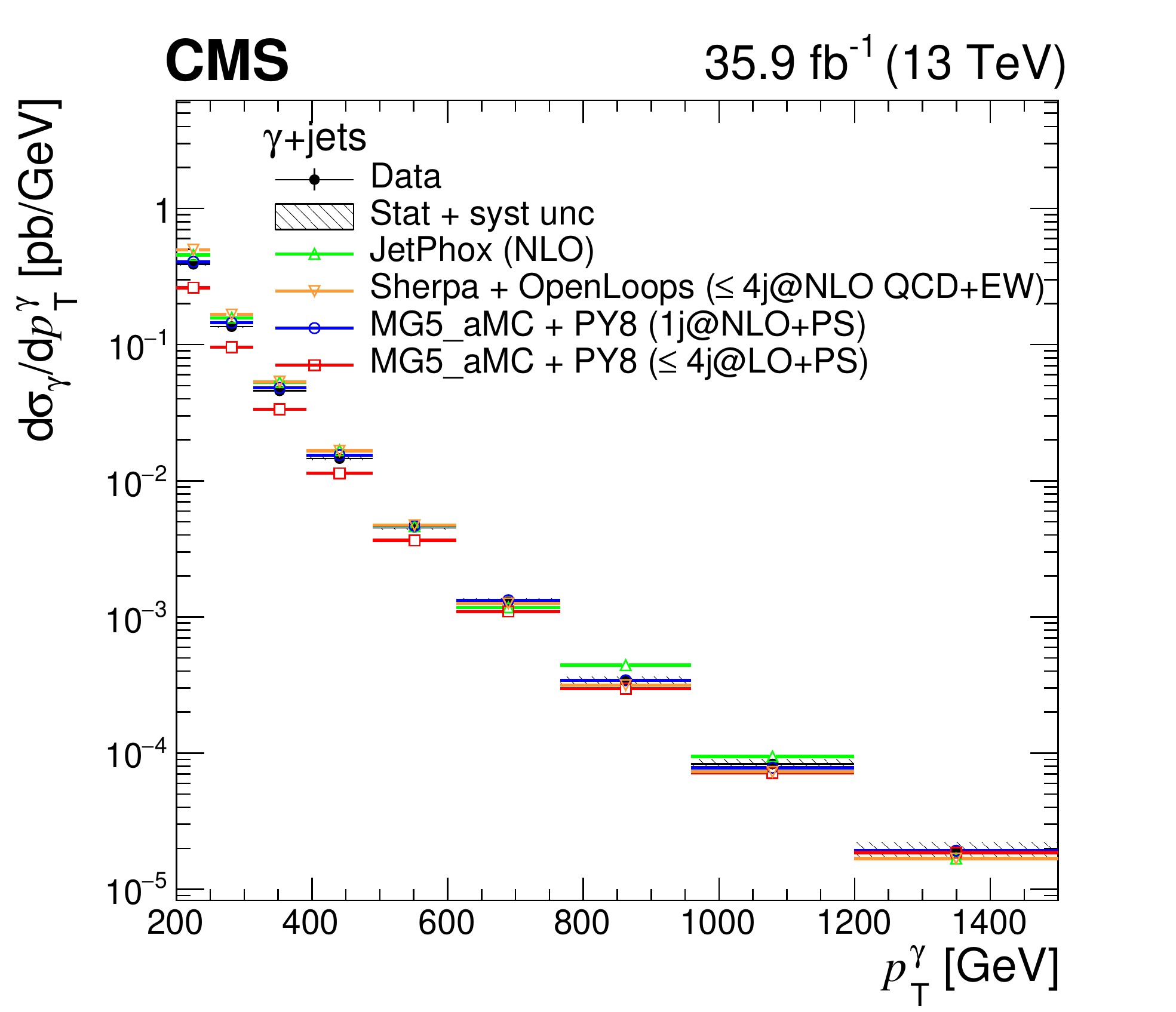}\\
  \includegraphics[width=0.49\textwidth]{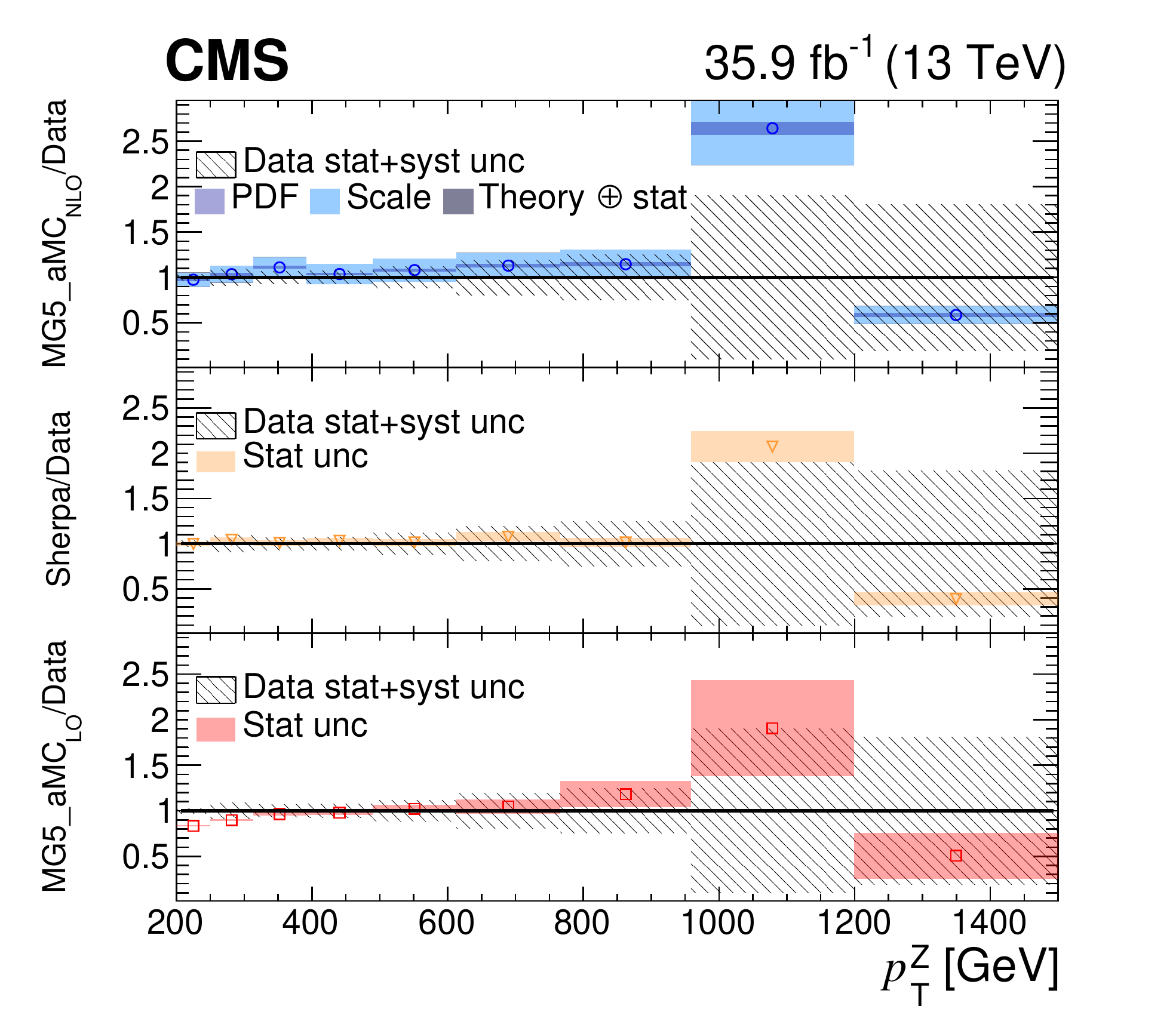}
  \includegraphics[width=0.49\textwidth]{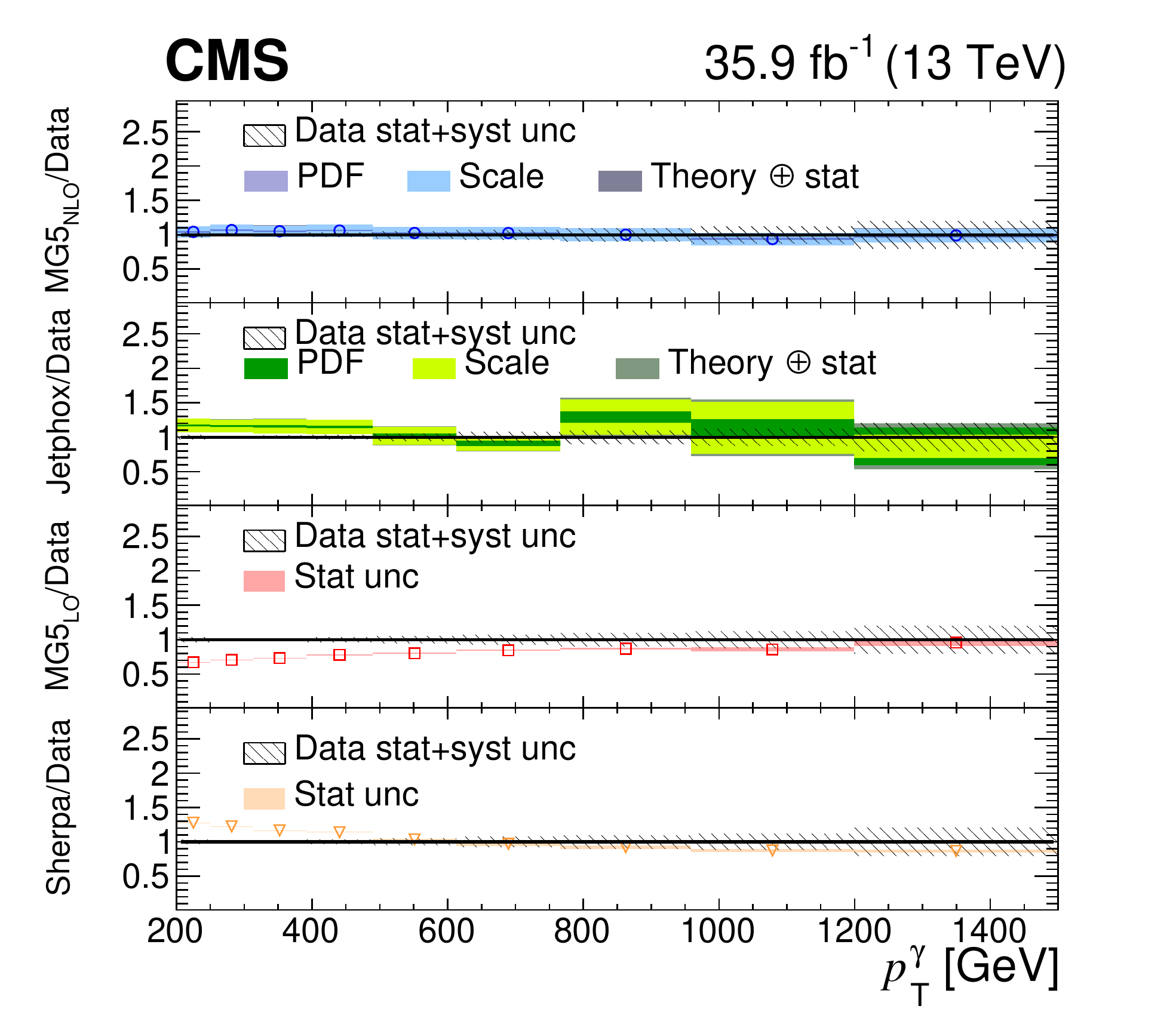}
\caption{Measured differential cross sections as a function of the boson \pt for \zj (\cmsLeft) and \gj (\cmsRight) and their comparisons with several theoretical predictions. The LO \MGvATNLO prediction for \zj has been normalized to the NNLO cross section (denoted by k$_{\textrm{NNLO}}$).
The vertical bars in the upper panels represent the statistical uncertainty in the measurement and the hatched band in the lower and upper panels is the sum in quadrature of the statistical and systematic uncertainty components in the measurement. The lower panels show the ratios of the theoretical predictions to the unfolded data. The shaded band in the LO \MGvATNLO and \SHERPA + \OPENLOOPS calculations shows the statistical uncertainty. The dark (light) shaded band on the NLO prediction from \MGvATNLO and the \JETPHOX prediction represents the PDF (scale) uncertainties, whereas the statistical uncertainties are barely visible.}
\label{fig:Zunfold}
\end{figure}

A comparison of the unfolded cross section of \zj events, as a function of the \PZ boson \pt, with several theoretical predictions is shown in Fig.~\ref{fig:Zunfold} left (upper and lower panels). The unfolded data are compared with the LO and NLO predictions from \MGvATNLO and the NLO QCD+EW prediction from \SHERPA + \OPENLOOPS. The predictions from \MGvATNLO at LO are normalized to the NNLO cross section from \FEWZ.
Statistical uncertainties associated with the MC are shown for the LO and NLO predictions in the lower panel. Additionally, the NLO prediction from \MGvATNLO is shown with the uncertainties from the variation in the PDFs, and in \mur and \muf. The PDF uncertainty is evaluated by taking the one sigma uncertainty band from the different PDF replicas and the scale uncertainty is evaluated by independently varying the scales up and down by a factor of two, with the condition that $0.5 < \muf/\mur < 2.0$, and taking the largest cross section variation as the uncertainty.

The statistical uncertainty in the data is shown with the error bars, and the combined statistical and systematic uncertainty is represented by the shaded band. Systematic uncertainties from the integrated luminosity and the muon selection efficiency determination dominate the low-\pt region, whereas the major source of the systematic uncertainty in the high-\pt region comes from the unfolding. 
The precision in the high-\pt region is limited by both the statistical uncertainties in the data and also by the limited size of the MC samples. 
The data show agreement within uncertainties with all predictions across almost the entire \pt range. A difference of 1.7 standard deviations is observed between the data and the prediction from \MGvATNLO in the 950--1200\GeV bin. In general, the NLO calculations describe the shape of the \PZ boson \pt distribution better than the LO calculation.

The distribution of the unfolded photon \pt is compared with theoretical predictions from \JETPHOX, \SHERPA + \OPENLOOPS, and \MGvATNLO in Fig.~\ref{fig:Zunfold} right (upper and lower panels).
The LO prediction from \MGvATNLO shows a 10--30\% disagreement in the shape of the photon \pt distribution, in particular for \pt values below $\approx 600\GeV$. The corresponding NLO calculation shows agreement within uncertainties across the full range of \pt. 
The \SHERPA + \OPENLOOPS calculation overpredicts the data by 20--30\% in the \pt region below 500\GeV and is consistent within uncertainties for the rest of the \pt range. 
The NLO prediction from \JETPHOX is shown with the uncertainties from the variation in PDFs and in \mur, \muf and $\mu_{\mathrm{f}}$. 
The prediction is mostly consistent with data within uncertainties with a general overprediction at the level of $\approx 20\%$ for $\pt < 500\GeV$. Since the experimental uncertainties are smaller than or comparable with the theoretical uncertainties for low and intermediate photon \pt, this analysis can be useful to constrain the proton PDF~\cite{Carminati:2012mm}. 

\begin{figure}[!ht]
\centering
\includegraphics[width=0.49\textwidth]{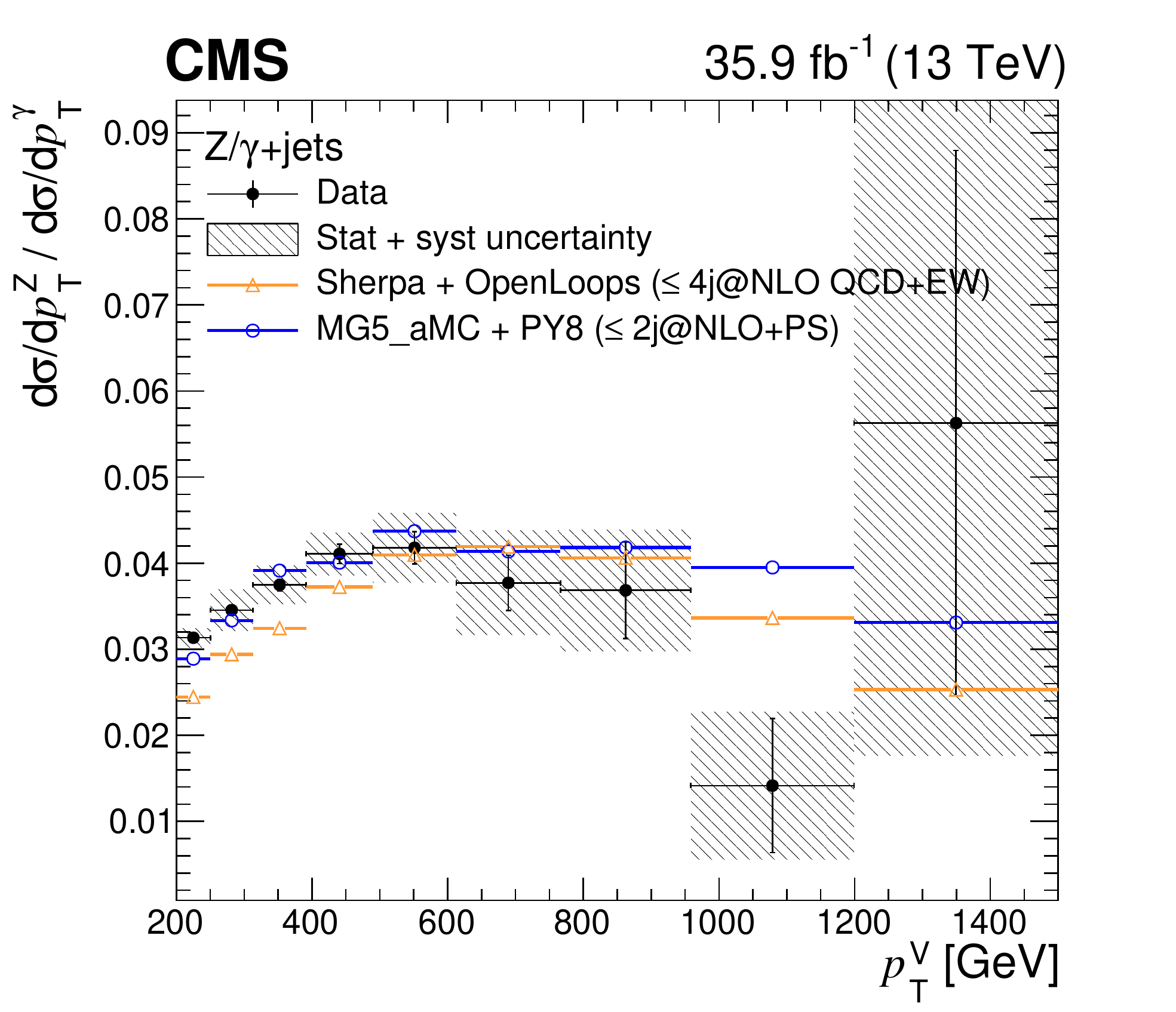}\\
\includegraphics[width=0.49\textwidth]{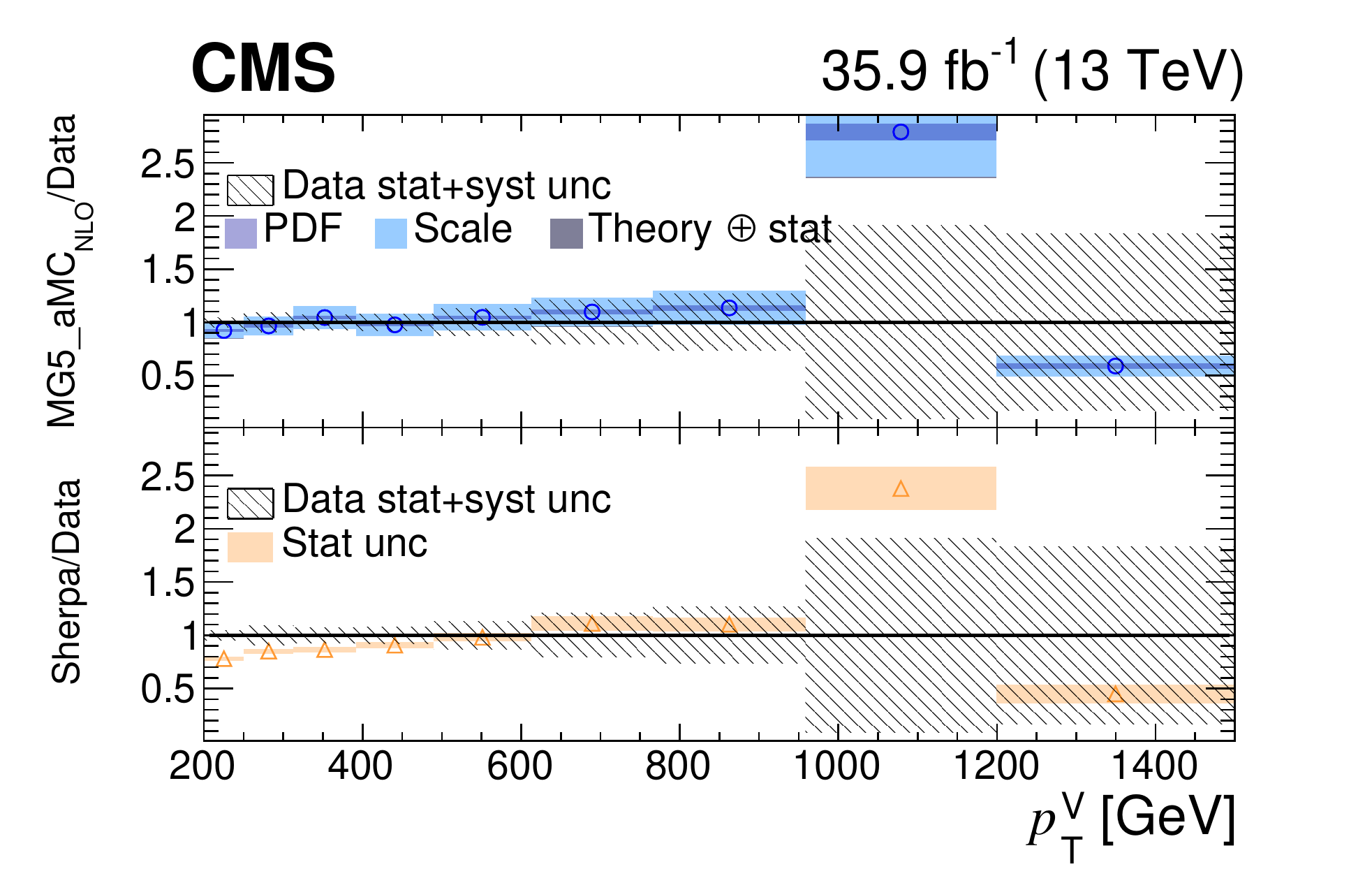}
\caption{Differential cross section ratio of \zj to \gj as a function of the vector boson (V) transverse momentum compared with the theoretical prediction from \MGvATNLO and \SHERPA + \OPENLOOPS. Only vector bosons produced centrally, with $\abs{y} < 1.4$, in association with one or more jets are considered. The lower panel shows the ratio of the theoretical prediction to the unfolded data. The vertical bars in the upper panel represent the statistical uncertainty in the measurement and the hatched band in the lower and upper panels is the sum in quadrature of the statistical and systematic uncertainty components in the measurement. The dark (light) shaded band on the NLO prediction from \MGvATNLO represents the PDF (scale) uncertainties, which are treated as uncorrelated between \zj and \gj, whereas the statistical uncertainties are barely visible. The shaded band on the \SHERPA + \OPENLOOPS calculation is the statistical uncertainty.}
\label{fig:ratio}
\end{figure}

The ratio of differential cross sections for the two processes, \zj over \gj, is shown in Fig.~\ref{fig:ratio} and compared with the theoretical prediction at NLO from \MGvATNLO and NLO QCD+EW from \SHERPA + \OPENLOOPS. The comparison with \MGvATNLO shows consistency within the uncertainties across the entire \pt range, whereas \SHERPA + \OPENLOOPS underpredicts the data by 10--20\% at low \pt, because of the overprediction in the photon \pt distribution, but is consistent with data within uncertainties for $\pt > 300\GeV$. 

\begin{figure}[!ht]
\centering
 \includegraphics[width=0.49\textwidth]{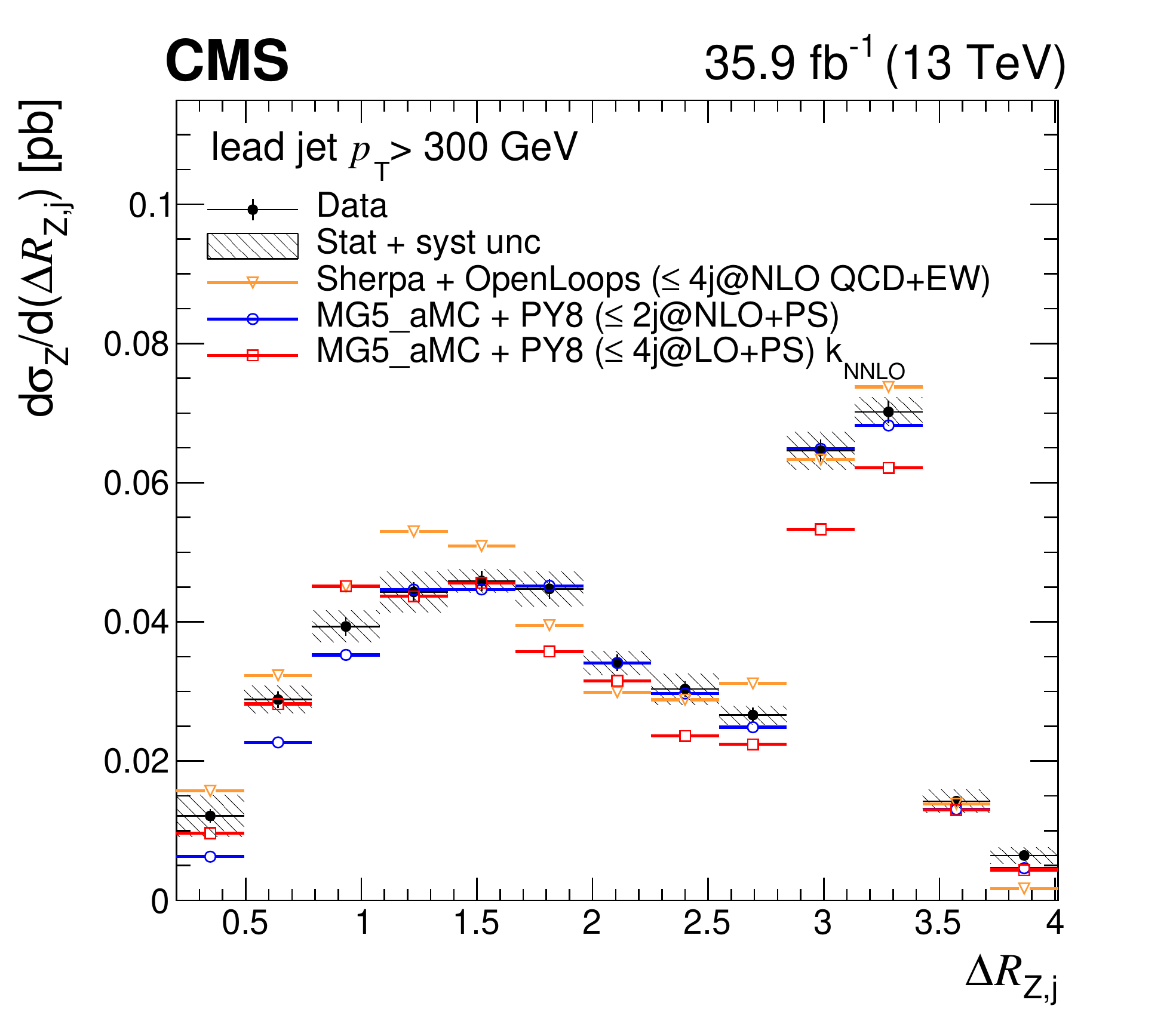}
 \includegraphics[width=0.49\textwidth]{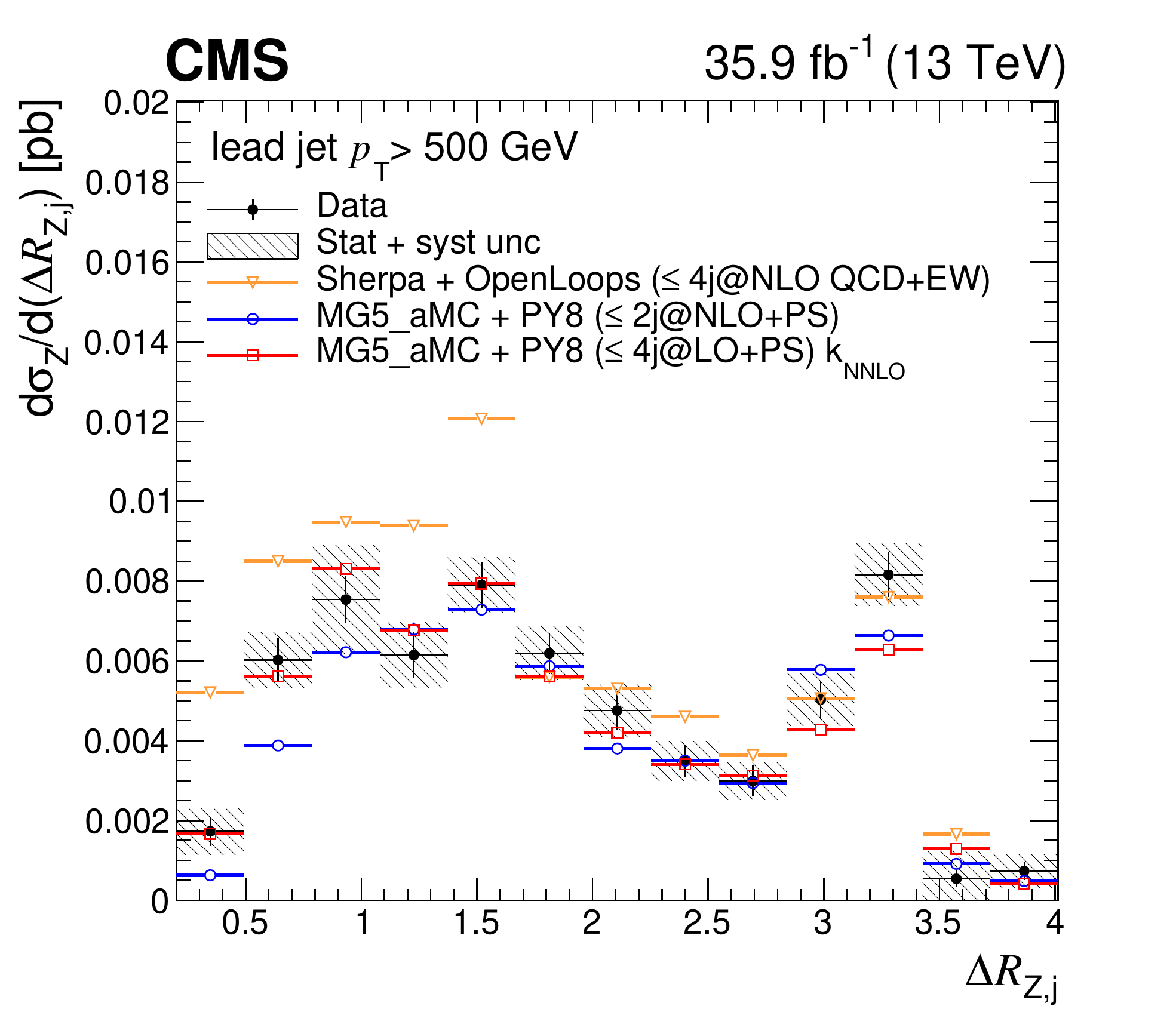}\\
 \includegraphics[width=0.49\textwidth]{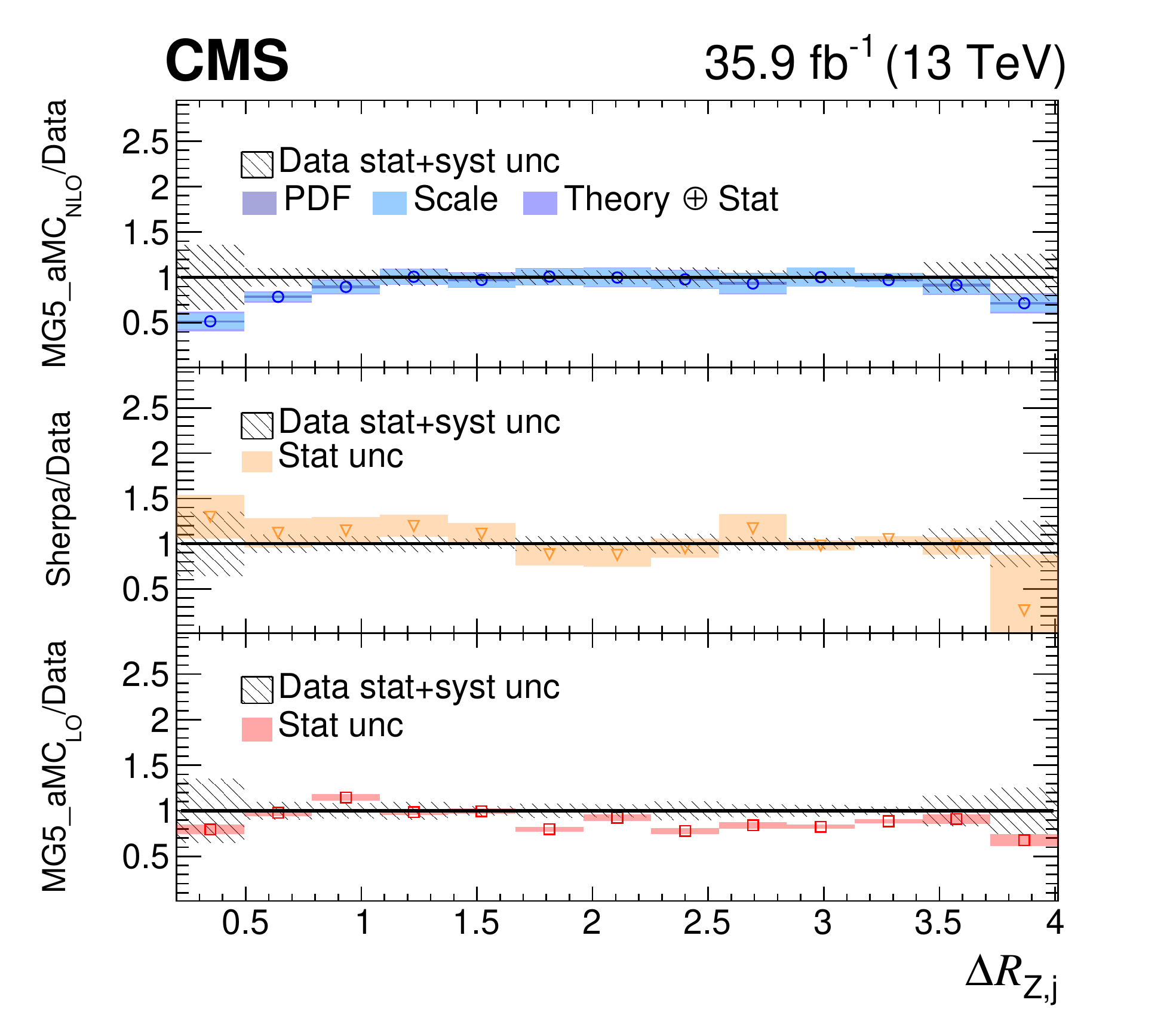}
 \includegraphics[width=0.49\textwidth]{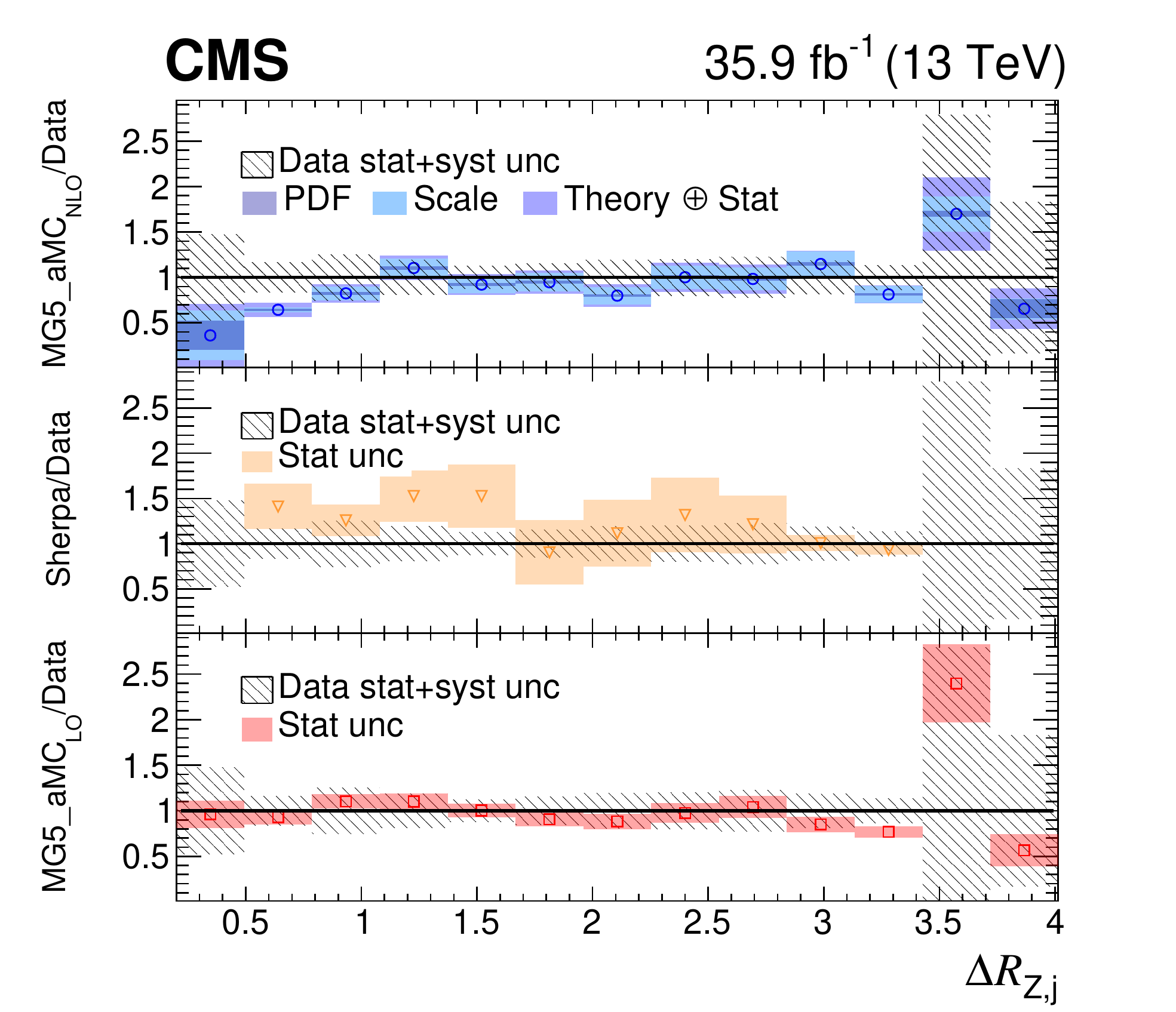}

\caption{Measured differential cross section of \zjets as a function of the angular separation between the \PZ boson and the closest jet, compared with theoretical predictions from \MGvATNLO and \SHERPA + \OPENLOOPS, where the leading jet \pt is above 300 (\cmsLeft) and 500 (\cmsRight)\GeV. The vertical bars in the upper panel represent the statistical uncertainty in the measurement and the hatched band in the lower and upper panels is the sum in quadrature of the statistical and systematic uncertainty components in the measurement. The lower panels show the ratio of the theoretical predictions to the unfolded data. The shaded band on the LO \MGvATNLO and \SHERPA + \OPENLOOPS calculations is the statistical uncertainty. The dark (light) shaded band on the NLO prediction from \MGvATNLO represents the PDF (scale) uncertainties.}
\label{fig:collinear}
\end{figure}

The unfolded distribution for the angular separation between the \PZ boson and the closest jet is shown in Fig.~\ref{fig:collinear} and compared with predictions from \MGvATNLO and \SHERPA + \OPENLOOPS. The peak around \mindr= 3.4 corresponds to the back-to-back production of a \PZ boson and a jet, whereas the region below $\mindr \approx 2.5$ is enhanced in the collinear emission of a \PZ boson close to a jet. 
The theoretical predictions are generally consistent with the data within the uncertainties for the case where the leading jet \pt is above 500\GeV. The LO prediction from \MGvATNLO underpredicts the data for $\mindr > 1.8$, whereas the NLO prediction is consistent within uncertainties for the bulk of the distribution with the largest discrepancy for \mindr below 0.8 for leading jet $\pt > 300\GeV$, the region dominated by collinear production. The \SHERPA + \OPENLOOPS prediction is typically higher than the data for the region below $\mindr \approx 1.8$, but has a large statistical uncertainty and is mostly consistent with the data within these uncertainties.

\section{Summary}

This paper presents measurements of standard model processes that probe regions of phase space characterized by the production of \zjets and \gjets at large boson transverse momentum (\pt), and of a \PZ boson in association with at least one very high \pt jet.

The measurements utilize data recorded with the CMS detector in $\Pp\Pp$ collisions at $\sqrt{s}$ = 13\TeV at the LHC that correspond to an integrated luminosity of 35.9\fbinv. Comparisons are made between the unfolded data and several theory predictions. 

The \zjets and \gjets cross sections as a function of boson \pt are presented for \pt above 200\GeV and compared with predictions from (i) the leading-order (LO) and next-to-leading-order (NLO) calculations from \MGvATNLO, and (ii) the NLO quantum chromodynamics and electroweak (QCD+EW) calculation from \SHERPA + \OPENLOOPS. In addition, the \gjets measurement is compared with NLO \JETPHOX predictions. The data are consistent with theory for both the $\gamma$ and \PZ boson final states, although in some regions of phase space a few tens of percent deviations are observed.  In general, the perturbative NLO corrections exhibit a better agreement with the measurements. 

{\tolerance=800 This is the first measurement at 13\TeV of the ratio of cross sections for \zjets to \gjets as a function of boson \pt. This ratio is compared with the NLO calculation from \MGvATNLO and the NLO QCD+EW prediction from \SHERPA + \OPENLOOPS; and it probes the region up to 1.5\TeV in boson \pt. The data are generally in agreement with theory within the uncertainties over the full range of boson \pt. This ratio provides an important theoretical input for the estimation, based on the \gjets process, of the \zvv background relevant in searches for new physics.\par}

The measurement of the emission of a \PZ boson collinear to a jet represents the first explicit study of this topology at the LHC. It is accessed through the production of a \PZ boson in association with at least one high-\pt jet ($>$300 or $>$500\GeV), and the differential cross section is measured as a function of the angular separation between the \PZ boson and the closest jet (\mindr). The unfolded data are compared with the LO and NLO calculations from \MGvATNLO, and the NLO QCD+EW prediction from \SHERPA + \OPENLOOPS. 
The NLO {\MADGRAPH} shows agreement over most of the measured distribution, but underpredicts it for the \mindr\ region below 0.8, which is dominated by events with the emission of a \PZ boson in close proximity to a jet. The prediction from \SHERPA is also generally consistent with the measurement.

The measurements presented in this paper will become increasingly important in current and future runs of the LHC, where the higher $\sqrt{s}$ and larger integrated luminosity will push the LHC program into new territory, necessitating an understanding of standard model processes in regions of previously unexplored phase space.

\clearpage

\begin{acknowledgments}
  We congratulate our colleagues in the CERN accelerator departments for the excellent performance of the LHC and thank the technical and administrative staffs at CERN and at other CMS institutes for their contributions to the success of the CMS effort. In addition, we gratefully acknowledge the computing centers and personnel of the Worldwide LHC Computing Grid and other centers for delivering so effectively the computing infrastructure essential to our analyses. Finally, we acknowledge the enduring support for the construction and operation of the LHC, the CMS detector, and the supporting computing infrastructure provided by the following funding agencies: BMBWF and FWF (Austria); FNRS and FWO (Belgium); CNPq, CAPES, FAPERJ, FAPERGS, and FAPESP (Brazil); MES (Bulgaria); CERN; CAS, MoST, and NSFC (China); COLCIENCIAS (Colombia); MSES and CSF (Croatia); RIF (Cyprus); SENESCYT (Ecuador); MoER, ERC PUT and ERDF (Estonia); Academy of Finland, MEC, and HIP (Finland); CEA and CNRS/IN2P3 (France); BMBF, DFG, and HGF (Germany); GSRT (Greece); NKFIA (Hungary); DAE and DST (India); IPM (Iran); SFI (Ireland); INFN (Italy); MSIP and NRF (Republic of Korea); MES (Latvia); LAS (Lithuania); MOE and UM (Malaysia); BUAP, CINVESTAV, CONACYT, LNS, SEP, and UASLP-FAI (Mexico); MOS (Montenegro); MBIE (New Zealand); PAEC (Pakistan); MSHE and NSC (Poland); FCT (Portugal); JINR (Dubna); MON, RosAtom, RAS, RFBR, and NRC KI (Russia); MESTD (Serbia); SEIDI, CPAN, PCTI, and FEDER (Spain); MOSTR (Sri Lanka); Swiss Funding Agencies (Switzerland); MST (Taipei); ThEPCenter, IPST, STAR, and NSTDA (Thailand); TUBITAK and TAEK (Turkey); NASU (Ukraine); STFC (United Kingdom); DOE and NSF (USA).
 
  \hyphenation{Rachada-pisek} Individuals have received support from the Marie-Curie program and the European Research Council and Horizon 2020 Grant, contract Nos.\ 675440, 724704, 752730, and 765710 (European Union); the Leventis Foundation; the Alfred P.\ Sloan Foundation; the Alexander von Humboldt Foundation; the Belgian Federal Science Policy Office; the Fonds pour la Formation \`a la Recherche dans l'Industrie et dans l'Agriculture (FRIA-Belgium); the Agentschap voor Innovatie door Wetenschap en Technologie (IWT-Belgium); the F.R.S.-FNRS and FWO (Belgium) under the ``Excellence of Science -- EOS" -- be.h project n.\ 30820817; the Beijing Municipal Science \& Technology Commission, No. Z191100007219010; the Ministry of Education, Youth and Sports (MEYS) of the Czech Republic; the Deutsche Forschungsgemeinschaft (DFG), under Germany's Excellence Strategy -- EXC 2121 ``Quantum Universe" -- 390833306, and under project number 400140256 - GRK2497; the Lend\"ulet (``Momentum") Program and the J\'anos Bolyai Research Scholarship of the Hungarian Academy of Sciences, the New National Excellence Program \'UNKP, the NKFIA research grants 123842, 123959, 124845, 124850, 125105, 128713, 128786, and 129058 (Hungary);the Council of Science and Industrial Research, India;  the Lebanese CNRS and the Lebanese University(Lebanon); the HOMING PLUS program of the Foundation for Polish Science, cofinanced from European Union, Regional Development Fund, the Mobility Plus program of the Ministry of Science and Higher Education, the National Science Center (Poland), contracts Harmonia 2014/14/M/ST2/00428, Opus 2014/13/B/ST2/02543, 2014/15/B/ST2/03998, and 2015/19/B/ST2/02861, Sonata-bis 2012/07/E/ST2/01406; the National Priorities Research Program by Qatar National Research Fund; the Ministry of Science and Higher Education, project no. 0723-2020-0041 (Russia); the Programa Estatal de Fomento de la Investigaci{\'o}n Cient{\'i}fica y T{\'e}cnica de Excelencia Mar\'{\i}a de Maeztu, grant MDM-2015-0509 and the Programa Severo Ochoa del Principado de Asturias; the Thalis and Aristeia programs cofinanced by EU-ESF and the Greek NSRF; the Rachadapisek Sompot Fund for Postdoctoral Fellowship, Chulalongkorn University and the Chulalongkorn Academic into Its 2nd Century Project Advancement Project (Thailand); the Kavli Foundation; the Nvidia Corporation; the SuperMicro Corporation; the Welch Foundation, contract C-1845; and the Weston Havens Foundation (USA).\end{acknowledgments}

\bibliography{auto_generated}

\cleardoublepage \appendix\section{The CMS Collaboration \label{app:collab}}\begin{sloppypar}\hyphenpenalty=5000\widowpenalty=500\clubpenalty=5000\vskip\cmsinstskip
\textbf{Yerevan Physics Institute, Yerevan, Armenia}\\*[0pt]
A.M.~Sirunyan$^{\textrm{\dag}}$, A.~Tumasyan
\vskip\cmsinstskip
\textbf{Institut f\"{u}r Hochenergiephysik, Wien, Austria}\\*[0pt]
W.~Adam, F.~Ambrogi, T.~Bergauer, M.~Dragicevic, J.~Er\"{o}, A.~Escalante~Del~Valle, R.~Fr\"{u}hwirth\cmsAuthorMark{1}, M.~Jeitler\cmsAuthorMark{1}, N.~Krammer, L.~Lechner, D.~Liko, T.~Madlener, I.~Mikulec, F.M.~Pitters, N.~Rad, J.~Schieck\cmsAuthorMark{1}, R.~Sch\"{o}fbeck, M.~Spanring, S.~Templ, W.~Waltenberger, C.-E.~Wulz\cmsAuthorMark{1}, M.~Zarucki
% ...
\vskip\cmsinstskip
\dag: Deceased\\
1:  Also at Vienna University of Technology, Vienna, Austria\\
% ...
\end{sloppypar}
\end{document}